\documentclass[epj]{svjour}
\usepackage{graphics,graphicx}
\usepackage{dcolumn}
\usepackage{bm}
\begin{document}

\title{Electron transport in a mesoscopic superconducting / 
ferromagnetic hybrid conductor}
\author{M. Giroud\inst{1}, K. Hasselbach\inst{1}, H.~Courtois\inst{1}, 
D. Mailly\inst{2} \and B.~Pannetier\inst{1}}
\institute{Centre de Recherches sur les Tr\`es 
Basses Temp\'eratures - C.N.R.S. associated to Universit\'e Joseph 
Fourier, 25 Av. des Martyrs, 38042 Grenoble, France \and Laboratoire de Photonique et de Nanostructures, Route de Nozay,
91460 Marcoussis, France}
\date{\today}
\abstract{
We present electrical transport experiments performed on submicron
hybrid devices made of a ferromagnetic conductor (Co) and a
superconducting (Al) electrode.  The sample was patterned in order to separate
the contributions of the Co conductor and of the Co-Al interface.  We
observed a strong influence of the Al electrode superconductivity on
the resistance of the Co conductor.  This effect is large
only when the interface is highly transparent.  We characterized
the dependence of the observed resistance decrease on temperature,
bias current and magnetic field.  As the differential resistance of
the ferromagnet exhibits a non-trivial asymmetry, we claim that the
magnetic domain structure plays an important role in the electron 
transport properties of superconducting / ferromagnetic conductors.
}
\authorrunning{M. Giroud et al.}
\titlerunning{Electron transport in a mesoscopic superconducting / 
ferromagnetic hybrid conductor}
\maketitle

\section{Introduction}

The question whether superconductivity can be induced in a
ferromagnetic metal is of fundamental and practical importance.  At
the junction of a Ferromagnetic metal (F) with a Superconductor (S),
the superconducting order parameter is predicted to oscillate and
decay rapidly in the ferromagnet as the distance to the F/S interface
increases.  The natural length scale is the exchange length
$L_{exch}=\sqrt{\hbar D/\mu_{B}H_{exch}}$, where $D$ is the electron
diffusion constant and $H_{exch}$ the exchange field expressed in
Tesla.  The latter expression holds in the dirty limit
$L_{exch}>l_{e}$, where $l_{e}$ is the elastic diffusion length. 
Physically, these effects occur because of the wave vector difference
between spin-up and spin-down electrons at the Fermi level
\cite{FFLO,Beasley}. $L_{exch}$ is the length over which the
two electrons of an Andreev pair get a
phase difference of $\pi$.  Oscillating behaviors were recently
detected in measurements of the density of states of a F/S junction
\cite{Kontos} and in the Josephson supercurrent of a S/F/S junction
\cite{Ryazanov}.  These experiments involved F layers with a thickness
of the order of the exchange diffusion length $L_{exch}$ which was
actually made rather large by choosing a ferromagnetic metal with a
small exchange field $H_{exch}$.  In conventional ferromagnetic
transition metals (Co which is used here, Ni, Fe, \ldots), the
exchange energy $\mu_{B}H_{exch}$ is large and greatly overcomes the
thermal energy $k_{B}T$ at cryogenic temperatures.  The corresponding
exchange diffusion length $L_{exch}$ is very small, of the order of a
few nanometers.

Many recent experiments involved mesoscopic F/S junctions with a
micron-scale ferromagnetic conductor made of Co or Ni.  Surprisingly, large
proximity effects were observed in the transport properties
\cite{PetrashovFerro,Giordano,Giroud98,PrlPetr} of the samples with a
transparent interface.  When compared to the conventional
superconducting proximity effect occurring in non-magnetic metals,
\cite{JLTPCourtois} this behavior suggests that the relevant
length scale is much larger than the expected coherence length
$L_{exch}$.  In the case of interfaces with an intermediate or low
transparency, the effect was shown to be restricted to the interface,
and was described within an extended BTK model \cite{Aumentado}. 
Nevertheless, it was surprising that the fit values of the parameter
$Z$, which is directly related to the interface transparency, varied so
little in comparison with the wide range of interface resistance.

Alternative explanations, ignoring the proximity superconductivity in
the ferromagnetic metal, have been proposed.  In the case of a
transparent interface, the spin accumulation at the F/S junction
\cite{vanWees,Falko,Belzig} could contribute significantly to the anomalous
transport properties, together with the Anisotropic Magneto-Resistance
(AMR) \cite{AMR}.  Spin accumulation arises near a F/S interface
because of the mismatch between the unpolarized pair current in S and
spin-polarized single electron current in F. An excess population of
minority-spin electrons therefore develops in F in the vicinity of the
interface over a length scale set by the spin-flip diffusion length
$L_{sf}$.  Another relevant mechanism could be the competition between
the fringe field of the ferromagnet and the diamagnetism of the
superconducting electrode.  This may result in an inhomogeneous
magnetic field distribution and affect locally AMR and/or the Hall
effect in the ferromagnet.  The AMR stems from spin-orbit coupling in
the ferromagnet and results in an anisotropy of the resistivity when
the angle between the local magnetization and the current flow
changes.  This enables the observation of magnetization reversal
processes in the resistance of small magnetic particles
\cite{APLAumentado}.

The microscopic mechanism inducing superconductivity in F close to the
S interface is the Andreev reflection where an incident electron is
reflected into a phase-correlated hole of the same spin.  This is
equivalent to creating an Andreev pair of electrons with opposite
spins.  Obviously, this process will be affected by the spin
polarization of the ferromagnetic metal and will even disappear in the
case of a fully-polarized metal \cite{Beenakker,Soulen,Upadhyaya}.  In
this respect, an open question is the role of the magnetic domain
structure.  Crossed Andreev reflections \cite{Feinberg} may appear at
the F-S interface close to a domain wall separating in F two magnetic
domains with opposite magnetization \cite{Melin}.  In the case of a
ferromagnet with an inhomogeneous magnetization, it was also proposed that
the spin-triplet component of the superconducting wave function
can have a strong amplitude \cite{Volkov,Kadigrobov}. 
Interestingly, this component should exhibit a slower spatial decay
than the usual singlet component.  This inhomogeneous
magnetization hypothesis is relevant because of both the shape
anisotropy of micro-fabricated structures and the effect of the
diamagnetism of the S electrode.

These open questions show the need for further investigation of
transport in a mesoscopic ferromagnetic conductor connected to a
superconductor.  Here, we report on transport measurements of
submicron Co-Al hybrid structures.  Compared to our previous
experiments \cite{Giroud98}, we modified both sample dimensions and
geometry to focus on transport properties in the ferromagnet itself,
near the superconducting contact.  We observed a large resistance drop
in samples with a highly transparent interface.  We studied the
dependence of this effect with the temperature, the magnetic field and
the bias current.  Our main result is the observation of a resistance
asymmetry between the two nominally identical branches of the Co wire.

\section{Samples description}

\subsection{Fabrication}

\begin{figure}
\includegraphics[width=5.3 cm]{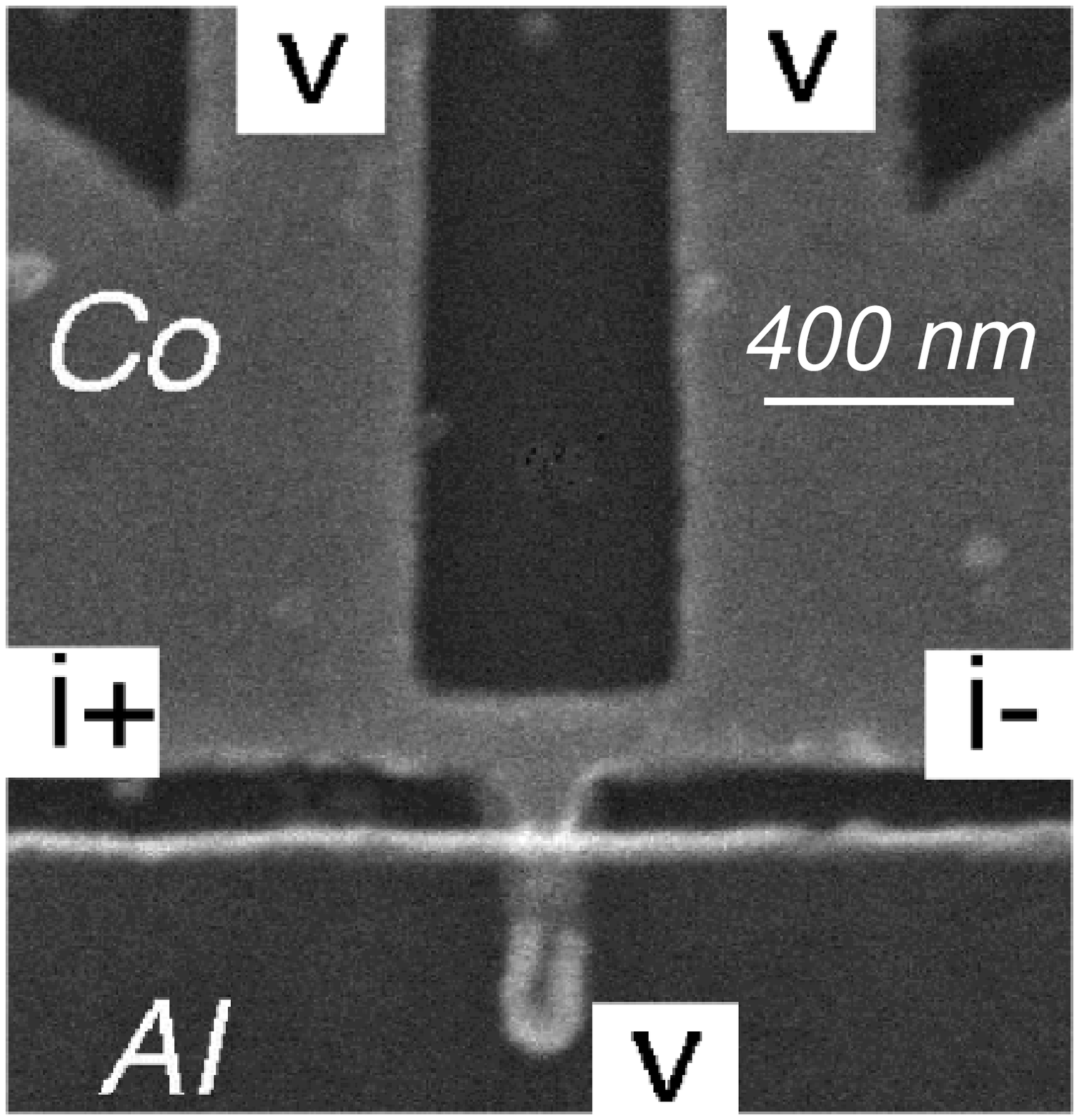}
\includegraphics[width=3.2 cm]{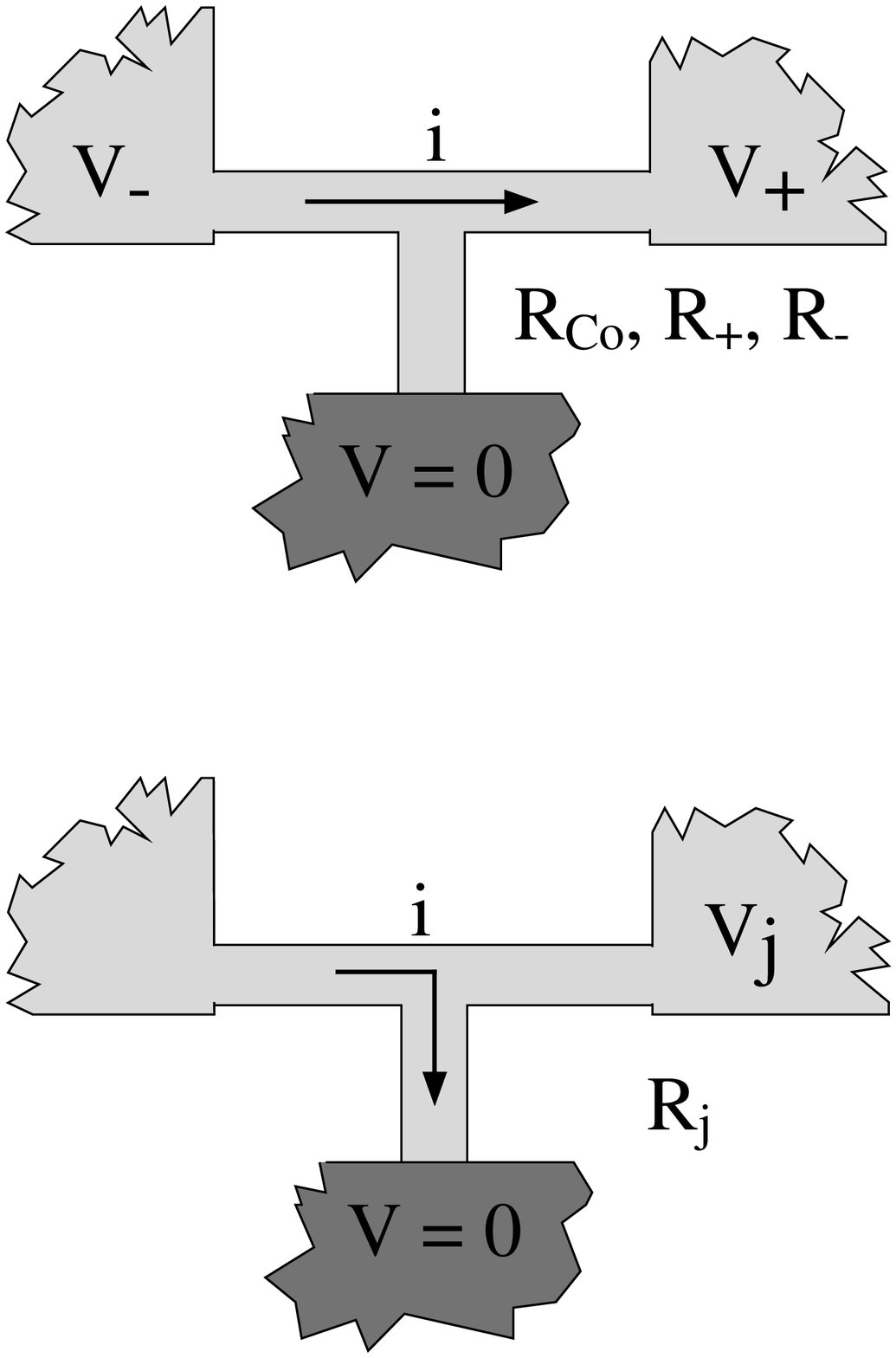}
\caption{\label{Photo} Left : Micrograph of a typical sample made of 
a small T-shaped conductor embedded between two Co reservoir (right 
and left) and one Al electrode (bottom).  The width and length of the small 
horizontal Co wire are respectively 120 nm and 400 nm.  Right : schematics of the measurement 
wiring.  The Co wire resistance $R_{Co}$ is measured by applying the 
bias current between two ferromagnetic pads and measuring $V_{+}-V_{-}$. The resistances $R_{+}$ and 
$R_{-}$ of the right and left arms are accessed by measuring only 
$V_{+}$ or $V_{-}$.  The junction resistance $R_{j}$ is measured by 
applying the current from one Co reservoir to the Al electrode and 
measuring the voltage $V_{j}$ of the opposite Co reservoir.}
\end{figure}

The samples geometry (see Fig.  \ref{Photo}) was designed for
measuring the ferromagnet resistance in proximity to the
superconducting contact, with zero net current through the interface. 
The Al contact is deposited on a lateral Co "finger", rather than
directly on top of the Co wire itself, in order to minimize spurious
current density redistribution effects when Al becomes superconducting
\cite{NSa2D,Belzig}.  As there are evidences that the superconducting
contact influences transport only in the vicinity of the interface,
the Co strip length was chosen as short as possible, namely 400 nm. 

The sample fabrication process was chosen as to keep F/S contact
resistances as small as possible.  We used a two-step lift-off process
with in-situ Ar ion etch.  Co was deposited first on the silicon
substrate, in order to avoid step edges which could modify
magnetization anisotropy and pin magnetic domain walls.  A 50 nm layer
of Co was e-beam evaporated at room temperature through a PMMA mask in
a base vacuum below $10^{-7}$ mbar.  An in-situ Ar ion milling of the
Co surface was performed just before the 100 nm Al layer evaporation
through the second PMMA mask.

\subsection{Characterization}

The deposition conditions together with the Co thickness are expected
to result in an in-plane magnetization, which was confirmed by the AMR
data.  This orientation makes it easier to induce magnetization
reversal under an applied magnetic field.  The expected typical domain
size is of the order of 100 nm, i.e. roughly comparable to the wire
width.  The Co resistivity was reproducibly high, in the 80 $\mu
\Omega$.cm range, whereas Al residual resistivity did not exceed 2
$\mu \Omega$.cm.  This corresponds to an electron diffusive mean free
path $l_{e}$ of 1 nm in Co and 20 nm in Al.  This also gives an
estimated coherence length for superconducting correlations $L_{exch}$
in Co of 3 nm.  The Al superconducting coherence length and London
penetration depth are of the order of 0.12 $\mu$m and 0.18 $\mu$m
respectively.  Overall, the Al resistive transition at $T_{c} \simeq$
1.3 K did not seem strongly affected by the proximity of the Co wire,
except for a depressed critical current, roughly 3 times lower than
for pure Al strips or microbridges of comparable dimensions.

We are mainly interested by the case of low resistance interfaces, as
it is presumably the only case where the superconductor has a strong
influence on the ferromagnet.  Interface resistance can be probed
using two probes on the Al electrode and two probes at both ends of
the Co strip.  Due to the sample geometry, we can only measure a
global resistance $R_{j}$ which is the interface resistance in series
with the Co lateral finger resistance and the Co spreading resistance. 
In the following, we consider experimental data collected on two
samples (A and B) with a low resistance interface.  The distance d
between the main Co wire and the Al contact, i.e. the length of the
lateral Co finger not covered by Al, was estimated from SEM
micrographs to d $\simeq$ 50 nm in sample A and d $\simeq$ 100 nm in
sample B. The global "junction" resistance $R_{j}$ was about 17
$\Omega$ in sample B. From the sample dimensions (see Fig. 
\ref{Photo}) and Co resistivity, we estimate that the Co finger itself
is the main contribution.  We can thus deduce that our interface
specific resistance is below 6 $m\Omega.\mu$m$^2$ in these samples
with a transparent interface.  For comparison, we also discuss data
from one sample (C) with a degraded interface.  The resistance $R_{j}$
is 103 $\Omega$, which leads to an estimated interface specific
resistance of 600 $m\Omega.\mu$m$^2$.

\section{Experimental results}

\subsection{Measurement procedure}

We studied the electron transport properties of several samples down
to a temperature of 30 mK. A magnetic field was applied in the sample
plane, either parallel or perpendicular to the current in the Co wire. 
An experimental run with a field perpendicular to the substrate plane
was also performed on some samples, to check that the Co AMR had the
behavior expected for in-plane magnetization.  In all our
measurements, a d.c. current bias plus an a.c. current modulation below
100 nA were applied through the same contacts.  Low-pass filters were
inserted on the cryostat feedthroughs, as well as a d.c. rejection
filter at the input of the lock-in amplifier.  Depending on the wiring
(see Fig.  \ref{Photo} right part), we investigated the resistance
$R_{Co}$ of the Co wire between the two reservoirs, that of one of the
halves of this wire $R_{+}$ or $R_{-}$, or the junction resistance
$R_{j}$.

\begin{figure}
\includegraphics{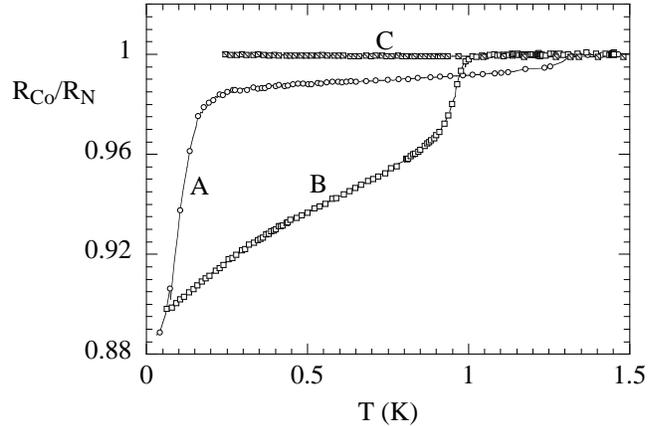}
\caption{\label{R(T)AB} Temperature dependence of the samples A, B and
C Co wire resistance ratio after zero field cooling.  The residual
resistance of the Co wire at 1.5 K just above Al superconducting
transition is 104 $\Omega$ for sample A, 111 $\Omega$ for sample B,
103.5 $\Omega$ for sample C.}
\end{figure}

\begin{figure}
\includegraphics{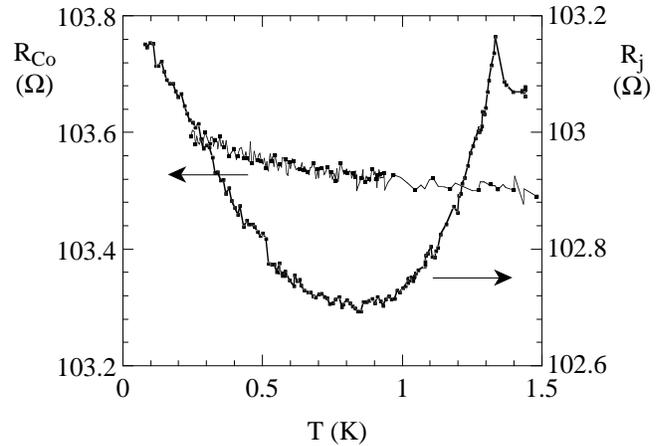}
\caption{\label{R(T)C} Temperature dependence of Co wire resistance
and Co-Al interface resistance in sample C. Note the difference 
in resistance change scale compared to Fig.  \ref{R(T)AB}.  }
\end{figure}

\subsection{Temperature dependence}

Fig.  \ref{R(T)AB} shows the temperature dependence of the Co
conductor in samples A, B and C. As the temperature is decreased below
the critical temperature of Al, the resistance of sample A first drops
by about 1 $\%$.  It drops again by about 10 $\%$ below 0.2 K.
Qualitatively similar results were obtained in other samples with a
low resistance Co-Al interface.  This is the case of sample B, which
was patterned on the same wafer than sample A, but then the
temperature of the large resistance drop is much higher whereas the
first resistance drop just below 1.3 K is not so clearly visible. 
Therefore the characteristic temperature of the resistance drop, as
well as the shape of the curve, appear to be sample-dependent.  We do
not yet know precisely which factors monitor this variation.  Let us
point out that the total resistance drop magnitude is nearly 12 $\%$
in samples A and B. This effect is not observed at large bias current.

In the case of sample C with a poor interface transparency, the
superconducting transition has almost no effect on the ferromagnet
resistance $R_{Co}$, as shown in Fig.  \ref{R(T)AB}.  It only weakly
affects the junction resistance $R_{j}$, which first decreases of
about 0.4 $\%$, and then increases on cooling down (see Fig. 
\ref{R(T)C}).

\subsection{Magnetoresistance}

\begin{figure}
\includegraphics{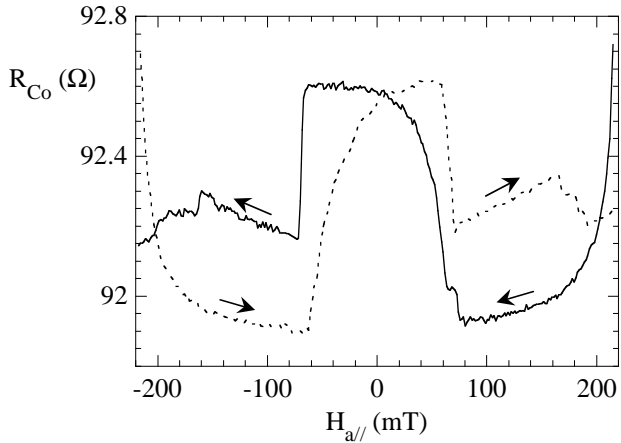}
\caption{\label{MR} Magneto-resistance of the Co wire in sample A. 
Magnetic field is applied parallel to the Co strip. The modulation 
current is 100 nA and the temperature 30 mK.}
\end{figure}

Fig.  \ref{MR} shows the magneto-resistance of sample A at low bias
current for a magnetic field applied in-plane along the Co wire, i.e.
parallel to the current path.  At low field ($H <$ 150 mT), the
magnetoresistance shows a small amplitude (less than 1 $\%$) and a
significant hysteresis.  These two features suggest that this
low-field behavior is due to Anisotropic Magneto-Resistance (AMR). 
The demagnetized multi-domain state reproducibly exhibits a larger
resistance than the higher field state.  The relatively sharp jumps at
$\pm$ 70 mT are the signature of the cobalt coercitive field
$H_{coer.}$.  In the high-field regime, we observe a large resistance
increase up to the normal state residual resistance.  This increase is
the signature of the Al transition to the normal state at the Al
superconducting critical field $H_{cS} \simeq$ 210 mT.

Measurements on sample C (not shown) showed a small 0.6 $\%$ positive
magneto-resistance when field and current are in-plane but
perpendicular to each other.  This anisotropy, as well as the
amplitude of the resistance jumps, are consistent with the effect of
the Co AMR in the case of an in-plane magnetization.  The small
resistance increase at very low temperature for sample C (see Fig. 
\ref{R(T)C}) can therefore be explained by a modification of the Co
AMR induced by Al diamagnetic shielding.

Let us note a significant difference with our previous experiments 
\cite{Giroud98} on 2 $\mu$m-long samples. In these long samples, 
magnetoresistance showed a smooth variation over a relatively broad 
field range. In the present short "T-shaped" samples, we reproducibly 
observe very well defined jumps at $\pm H_{coer.}$.  This strongly suggests 
that our short wires contain a small number of domain walls which 
depin at this field, whereas long samples contain a larger number of 
domains walls, which will rotate or move along the sample in a 
broader field range.

\subsection{Differential resistance}

\begin{figure}
\includegraphics{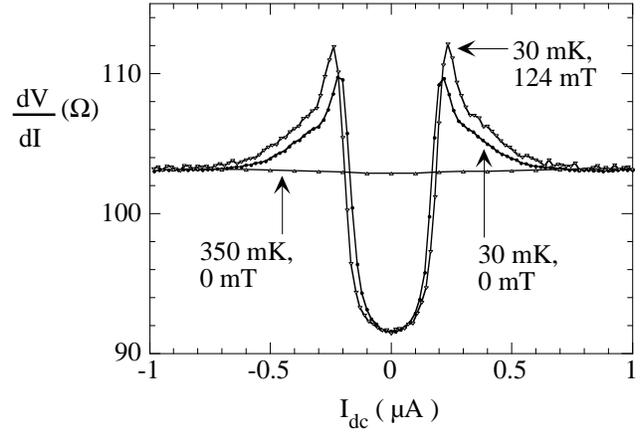}
\caption{\label{ResDif} Differential resistance dV/dI of sample A Co 
wire in various conditions : zero magnetic field, temperatures T = 30 mK and 
T = 350 mK ; magnetic field H = 124 mT, temperature T = 30 mK.}
\end{figure}

The differential resistance was investigated by superposing a d.c.
current bias current to the a.c. modulation.  Fig.  \ref{ResDif}
displays the total differential resistance measured in a standard four
probes configuration between the "+" and the "-" ends of the Co wire. 
In sample A, we observe a resistance dip for current bias below 0.2
$\mu$A which mimics the resistance drop observed below 0.2 K. At this
point, the voltage across sample A is about 20 $\mu$V, roughly a
factor 10 below the expected Al gap.  The high-bias resistance value
coincides with the value obtained at high temperature, above the
critical temperature of Al.  The relative amplitude of the variation
is in the 10 $\%$ range.  A similar behavior was reproducibly observed
in sample B and several other samples.  The current range and the
profile of the resistance dip appear to scale with the characteristic
temperature of the resistance drop.  No variation of the junction
differential resistance was observed at low bias except for
temperatures close to $T_{c}$.

Because of the relatively high resistivity of our Co film, we have to
be careful about heating effects.  At the lowest temperature (30 mK),
we expect that the heat flow is essentially evacuated along the Co
strip.  At a temperature of 0.1 K, the thermal conductance of the Co
strip, Co/Si interface, and Al strip, should be respectively about
100, 50 and 10 pW/K. The Joule power dissipated by the Co strip is
only 4 pW at the 0.2 $\mu$A current bias required to suppress the
resistance decrease.  This means that at this bias the sample A cannot
be heated above 0.1 K. We conclude that Joule heating is not
sufficient to explain the differential resistance variation.

We also applied a $\pm$ 124 mT field along the Co strip, parallel to
the current path.  The value $\pm$ 124 mT was chosen as to be above
the Co coercitive field $H_{coer.} \simeq$ 70 mT and below the Al superconducting
critical field $H_{cS} \simeq$ 210 mT. The Co conductor was therefore presumably
close to magnetization saturation, although it may not yet be single
domain.  The current dependence of $R_{Co}$ shown in Fig. 
\ref{ResDif} only shows a small difference compared to the zero field
case.  This difference is of the same order than the Co
magnetoresistance jump at $H_{coer.}$, i.e. close to 1$\%$.
 
\subsection{Resistance asymmetry}

\begin{figure}
\includegraphics{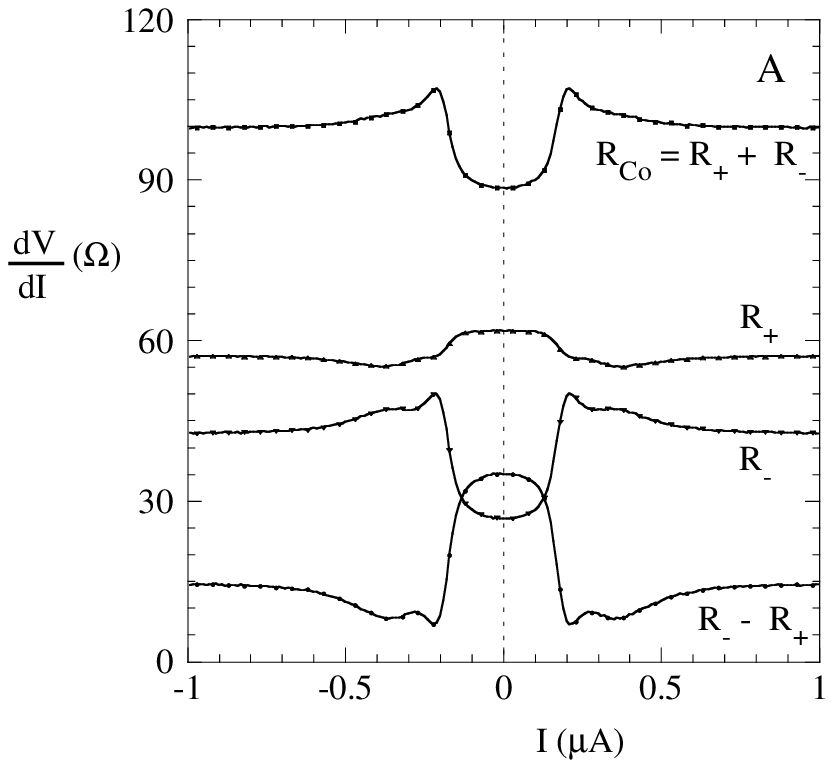}
\includegraphics{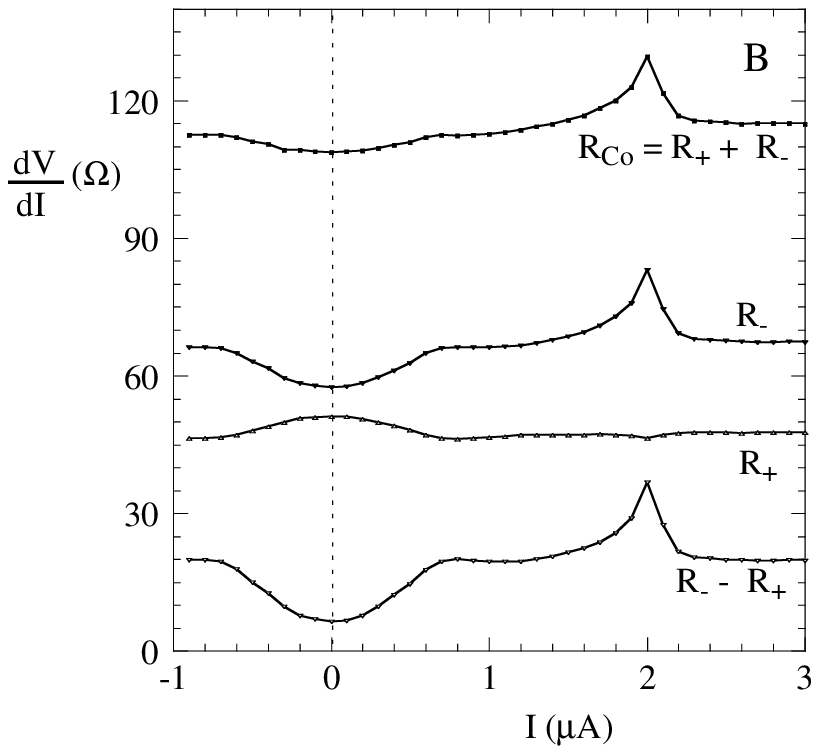}
\caption{\label{Parties} Current bias dependence of the differential  
resistance of "+" and "-" parts of the Co wire in samples A (top) at 
T = 30 mK, and B (bottom) at T = 700 mK. In both cases, the sum of the 
two resistances $R_+$ and $R_-$ matches the measured total resistance 
R.}
\end{figure}

We now come to our main result.  Figure \ref{Parties} top part shows
sample A differential resistances of the left and right parts of the Co
wire ($R_{+}$ and $R_{-}$), as well as the total resistance 
$R_{Co}=R_{-}+R_{+}$ and the resistance difference $R_{-}-R_{+}$. 
The current path between the two ends of the Co strip remained
unchanged, so that the net current through the interface is always
zero.  $R_+$ and $R_-$ represents the differential resistance measured
respectively with voltage probes at the "+" Co end and the Al contact,
or between the Al contact and the "-" Co end (see Fig.  \ref{Photo}).

We observe a non-trivial asymmetry between $R_+$ and $R_-$ in the low
bias regime.  In the current range where Co resistance is depressed by
the superconducting contact, $R_-$ is lower and $R_+$ higher than its
respective high-bias value.  The relative variations $\Delta R_- /R_-$
and $\Delta R_+ / R_+$ are significantly stronger (up to 35 $\%$) than
the total Co wire variation $\Delta R_{Co}/R_{Co}$ (about 12 $\%$). 
At high bias, the difference between the values of $R_{+}$ and $R_{-}$
may be explained by the inhomogeneities in the sample cristalline
micro-structure.  Let us note that the differential resistance remains
symmetric with respect to the d.c. current bias, i.e. a current
reversal produces a voltage sign reversal.  Within a small
experimental error, the sum of $R_+$ and $R_-$ is always equal to the
total differential resistance $R_{Co}$.  Similar results were
reproducibly observed on several samples, including sample B (Fig. 
\ref{Parties} bottom).

\section{Discussion of the results}

First, let us comment on a difference between these results and our
previous experiments \cite{Giroud98}: we no longer observed a
re-entrance of the metallic resistance.  Here, the estimated Thouless
energy of the Co conductor is $\epsilon_{c}=\hbar D/L^2 = 0.1K$, and a
resistance minimum should have been observed close to 0.5 K, which is
not the case.  This contradicts the interpretation we proposed
earlier, namely the occurrence of a re-entrant proximity effect in the
resistance, similarly to the non-ferromagnetic metal case
\cite{JLTPCourtois}.  We come to the conclusion that the resistance
minimum in our previous experiments may rather result from the
competition between two opposite mechanisms: a resistance drop
induced by the superconducting contact which we observe much more
clearly in our shorter new samples, and a resistance upturn of
different origin.  Spin accumulation effects constitute an obvious
candidate \cite{Belzig}, but Anisotropic Magneto Resistance (AMR)
cannot be excluded.

The first question we have to care about is whether the resistance
drop is indeed occurring in the ferromagnet, or is merely a
consequence of current redistribution in the superconducting
short-circuit \cite{NSa2D}.  It is important to note that in our
samples, the sheet resistance of Al above $T_{c}$ is only 0.2 $\Omega$
per square, which is much smaller than the sheet resistance of Co
($\simeq$ 20 $\Omega$ per square).  Even normal Al acts as a shunt. 
One can model our two-dimensionnal sample with an array of resistances
of the order of the Co and Al sheet resistances.  The result of this
analysis is that the resistance drop expected at Al superconducting
transition should not exceed the Al sheet resistance (0.2 $\Omega$). 
This is clearly much smaller than the experimental resistance drop,
which exceeds 10 $\Omega$.  It is also known that the AMR is of the
order of only 1 to 2$\%$ in ferromagnetic 3d transition metals such as
cobalt.  We indeed observe this AMR effect in the low-field
magnetoresistance (Fig.  \ref{MR}).  Thus we are confident that the
resistance drop observed in Co, about 10 $\%$ above the residual
normal state case, is neither a trivial current redistribution effect
in the superconducting short-circuit nor a simple AMR effect.  The
behavior of otherwise identical samples with respectively low or high
junction resistance is so clearly distinct that we can conclude that
interface transparency plays a key role.  The comparison between Fig. 
\ref{R(T)AB} and \ref{R(T)C} even shows that the sign of the 
resistance variation
can be reversed, as was observed in Ref.  \cite{PrlPetr}.  In the case
of a weakly transparent interface, our results are also compatible
with the conclusion of Ref.  \cite{Aumentado} that no proximity effect
appears in the bulk of the ferromagnet.  A significant influence from
the superconductor on the ferromagnet occurs when the interface is
transparent enough, and only in this case.

The differential resistance asymmetry when changing the voltage probes
configuration is somewhat surprising since the sample were fabricated
as symmetric.  We have to consider physical phenomena which may
contribute differently to the resistances $R_+$ and $R_-$ of the two
sample halves.  From the sample geometry shown in Fig.  \ref{Photo},
we see that $R_+$ and $R_-$ may include a Hall effect contribution,
related to the local field in the ferromagnet or to its magnetization
(anomalous Hall effect) over the Co transverse dimensions (wire width
plus finger length).  The Hall voltage will contribute with opposite
sign to $R_+$ and $R_-$, but should not contribute to the total
resistance $R_{Co}$, as in this case voltage probes are aligned along
the current lines.  A d.c. current bias as low as 0.2 $\mu$A cannot
significantly affect the magnetization and the domains structure
\cite{tsoi,wegrowe01}, so that the Hall resistance should remain
constant in the experimental bias range.  Moreover, we observed that
samples with a high resistance interface do not exhibit such a
significant resistance asymmetry, although their magnetization and
coercitive field are not significantly different.  Therefore, the
differential resistances and their asymmetry cannot be explained by the
classical anomalous Hall effect in the ferromagnet, but are related to
superconductivity.

In the interface region, electrons diffuse between Co and Al.  In the
case of a positive d.c. bias electrons coming from the "+" side have a
higher energy than those going on the "-" side.  An out-of-equilibrium
energy distribution will develop near the interface even with a zero
net current \cite{JLTPCourtois}.  Because of the mismatch between
spin-polarized current in F and unpolarized pair current in S, any
current between S and F will be also associated to a non-equilibrium
spin polarization, in a distance range from the interface determined
by the spin flip diffusion length.  This spin polarization is
energy-dependent, should be maximum at zero bias and vanish above the
superconducting gap.  With this simple picture in mind, one might
understand that $R_+$(I) and $R_-$(I) may be different at low bias. 
Nevertheless, we expect that $R_-$(-I) should behave as $R_+$(I) if
the sample is symmetric.  This is clearly not the case, one exhibiting
a maximum at zero bias and the other one a minimum.

Therefore, the difference between $R_+$ and $R_-$ must stem from a
physical asymmetry in the sample itself, the most obviously possible
one being the magnetic domains structure in an otherwise symmetric
geometry.  Since the T shape results in a complicated shape
anisotropy, it is very likely that a non-symmetric magnetic domain
structure is present in the central region of our sample.  The
amplitude and the sign of this asymmetry will be of course
sample-dependent.  If we assume that there is only a small number of
domains in our short samples, and if the spin flip diffusion length is
not much smaller than domain size (e.g. in the 200 nm range), the
chemical potential drop may be different for electrons travelling from
one Co contact (or the other) to the Al electrode, depending on the
magnetic domain structure in the $R_+$ or $R_-$ section of the cobalt
wire.  Nevertheless, the actual effect of the superconductivity on the
electron transport in Co and the physical origin of the resistance
drop remains undetermined.  It could be a long-range superconducting
proximity effect like the predicted long-range triplet component
\cite{Volkov,Kadigrobov}, spin accumulation effect in close
relation with the sample geometry \cite{Falko,Belzig}, or Andreev
reflections of electrons of opposite spins in adjacent ferromagnetic
domains of different magnetization \cite{Feinberg,Melin}.

\section{Conclusion}

In this work, we brought new experimental evidence for large
resistance decrease in hybrid Ferromagnetic / Superconducting devices. 
This effect is clearly distinct from the Anisotropic MagnetoResistance
(AMR) of smaller amplitude.  A high interface transparency was found
to be necessary for observing large effects.  We suggest the relevance
of the magnetic domain structure in the transport properties in the
vicinity of a F/S contact.  Further work on transport properties of
F/S junctions with a high control of the magnetic domain
structure is in progress .

We thank W. Belzig, R. M\'elin and M. Viret for discussions and C.
Veauvy for help in cryogeny.  We acknowledge discussions within the
"Dynamics of Superconducting Nanocircuits" TMR network and the
"Mesoscopic Electronics" COST P5 action.  This work is supported by the
French Ministry of Education and Research under an ACI contract.

\end{document}